# Spectroscopic size and thickness metrics for liquid-exfoliated h-BN

Aideen Griffin,[1] Brian Cunningham,[2] Declan Scullion,[2] Tian Tian,[3] Chih-Jen Shih,[3] Myrta Gruening,[2] Andrew Harvey,[1] John Donegan,[1] Elton J. G. Santos,[2] Claudia Backes,[4] Jonathan N. Coleman[1*]

[1]*School of Physics and CRANN & AMBER Research Centres, Trinity College Dublin, Dublin 2, Ireland*

[2]*School of Mathematics and Physics, Queen's University Belfast, Belfast, BT71NN, United Kingdom*

[3]*Institute for Chemical and Bioengineering, ETH Zürich, 8093 Zürich, Switzerland*

[4]*Chair of Applied Physical Chemistry, Ruprecht-Karls University Heidelberg, Im Neuenheimer Feld 253, 69120 Heidelberg, Germany*

*colemaj@tcd.ie

**Abstract**: For many 2D materials, optical and Raman spectra are richly structured, and convey information on a range of parameters including nanosheet size and defect content. By contrast, the equivalent spectra for h-BN are relatively simple, with both the absorption and Raman spectra consisting of a single feature each, disclosing relatively little information. Here, the ability to size-select liquid-exfoliated h-BN nanosheets has allowed us to comprehensively study the dependence of h-BN optical spectra on nanosheet dimensions. We find the optical extinction coefficient spectrum to vary systematically with nanosheet lateral size due to the presence of light scattering. Conversely, once light scattering has been decoupled to give the optical absorbance spectra, we find the size dependence to be mostly removed save for a weak but well-defined variation in energy of peak absorbance with nanosheet thickness. This finding is corroborated by our *ab initio GW* and Bethe Salpeter equation calculations, which include electron correlations and quasiparticle self-consistency (QS*GW*). In addition, while we find the position of the sole h-BN Raman line to be invariant with nanosheet dimensions, the linewidth appears to vary weakly with nanosheet thickness. These size-dependent spectroscopic properties can be used as metrics to estimate nanosheet thickness from spectroscopic data.

**Introduction**

Hexagonal boron nitride (h-BN) is a layered material which is structurally analogous to graphite.[1] Its physical properties resemble graphite in a number of ways, for example in its high chemical stability, its large thermal conductivity and near superlative mechanical properties. However, it is electrically very different to graphite, displaying a large bandgap (5.5-6 eV) and negligible electrical conductivity.

Also like graphite,[2] h-BN can be produced in a 2-dimensional (2D) form by direct growth[3] as well as by mechanical[4] and liquid phase exfoliation.[5-6] The exfoliated material retains the properties of layered h-BN but in an ultra-thin, extremely flat morphology. This has resulted in 2D h-BN being deployed in a range of applications. For example, due to its high bandgap and extreme flatness, grown or mechanically exfoliated h-BN is widely used as a substrate or encapsulating material for electronic devices based on other 2D materials such as graphene or $MoS_2$.[7-10] Alternatively, liquid-exfoliated h-BN nanosheets (which tend to be a few layers thick and 100s of nm in length) have been used in a range of applications from reinforcing[6] or gas-barrier[11] fillers in polymer-based composites to thermally conductive inclusions[12] in oils to dielectric materials in electronic devices[13-15] and electrochemical separators in electrolytically gated transistors.[16]

As with other 2D materials, the utility of h-BN in applications increases the importance of our ability to characterize it. As with all 2D materials, basic characterization to measure nanosheet size and thickness can be performed by transmission electron microscopy and atomic force microscopy. However, statistical analysis of individual nanosheet measurements using these techniques is time consuming and tedious. In contrast, optical spectroscopy generally probes the ensemble and provides averaged information. However, compared to other 2D nanomaterials optical spectroscopic characterization of h-BN has yielded much less information. For example, while $MoS_2$ and $WS_2$,[17-18] and to a lesser extent graphene,[19] have information-rich optical absorption spectra which allow estimation of nanosheet size and thickness, the absorption spectrum of h-BN appears to be information-poor, displaying few features beyond a bandedge around 6 eV. Similarly, while the Raman spectra of $MoS_2$ and graphene yield information about nanosheet dimensions[19-20] and defect content,[21] the h-BN Raman spectrum contains a single line,[1] the properties of which have not been concisely linked to any physical properties of the nanosheets. Although cathodoluminesce can give information about nanosheet thickness,[22] these measurements are neither straightforward nor widely accessible.

Here we shown that the absorption and Raman spectra of liquid-exfoliated BN-nanosheets are not as bereft of information as has been previously thought. By performing optical characterization of fractions of size-selected, liquid-exfoliated nanosheets, we show that the extinction spectra are influenced by nanosheet lateral size while the nanosheet thickness can be extracted from either the absorption or Raman spectra.

**Results and Discussion**

*Size selection of BN*

Liquid phase exfoliation is a versatile nanosheet production method which exfoliates layered crystals down to few-layered nanosheets in appropriate stabilizing liquids.[23-24] It has been applied to a range of layered crystals including graphite, h-BN[5-6, 25-29] and $MoS_2$ and tends to yield polydisperse samples of nanosheets with broad lateral size (~100-1000 nm) and thickness (~1-20 layers) distributions.[23, 30-31] As a result, centrifugation-based size selection is required to enable any study where well-defined sizes are required. Such techniques range from density gradient ultracentrifugation,[30] which gives fine size control at low yield, to liquid cascade centrifugation (LCC),[18] which gives coarser size control at considerably higher yield. Here we used LCC to size-select an as-prepared dispersion of BN nanosheets stabilized in an aqueous sodium cholate solution (see Methods for exfoliation protocol and cascade details), yielding fractions containing nanosheets of different lateral sizes and thicknesses.

In LCC, a dispersion is subjected to repeated centrifugation steps with successively increasing centrifugal accelerations (expressed as relative centrifugal field, *RCF*, in units of the earth's gravitational field, *g*). After each centrifugation step, supernatant and sediment are separated, the sediments are collected for analysis, while the supernatant is centrifuged at higher centrifugal acceleration.[18] The sediments collected at low centrifugal acceleration contain large/thick nanosheets, while the fractions collected at higher centrifugal acceleration contain smaller and smaller nanosheets. This technique has a number of advantages; notably that collecting the product as a sediment allows redispersion into a range of liquid environments, simultaneously allowing solvent exchange and concentration increase. In addition, very little material is wasted with up to 95% of exfoliated product distributed among the fractions.[18]

We label samples using the lower and upper centrifugation rates used in the preparation of the fraction. For example, if the supernatant produced after centrifugation with *RCF*=5,000×*g*-force (5k-*g*) is then centrifuged at 10,000 *g* and the sediment collected after this step, we refer to the sample as 5-10 k-*g*.

Atomic Force Microscopy (AFM) was used to statistically analyze the nanosheet dimensions for each fraction with representative images displayed in Figure 1 A. In each dispersion, 200-350 nanosheets were measured, and their length (longest dimension), width (dimension perpendicular to length) and thickness recorded. The nanosheet length data were plotted as histograms with examples of the 0.4-1 k-*g* and 10-22 k-*g* fractions shown in Figure 1B. For each fraction, the nanosheet length follows a lognormal statistical distribution with smaller sizes obtained for increasing centrifugation speeds as expected. Additional histograms are given in the SI (Figure S1). The mean nanosheet length is plotted *versus* the central *g*-value (midpoint of high and low *g*-values used in the size selection) in figure 1C. Experimentally, we found a roughly power-law decay of <L> with central *g*-value with an exponent close to -0.5, similarly to other liquid-exfoliated 2D materials.[18]

Some care must be taken when analyzing the statistical nanosheet-height data. This is because the apparent AFM height of liquid-exfoliated nanosheets is typically larger than the theoretical thickness of the nanosheets due to adsorbed/intercalated water and surfactant. Similar to previous reports,[17, 32-33] we use step-height analysis to determine the apparent thickness of a single monolayer by measuring the height of the terraces of partially exfoliated nanosheets (Figure 1 D,E). Similar steps are then grouped and a plot of the mean step height versus step height group number (Figure 1, F) leads to an apparent monolayer height of 0.99±0.01 nm, similar to the step height of 0.9 nm previously found for graphene.[32] Using this information we can determine the number of layers, N, of the nanosheets allowing the construction of histograms for each size-selected fraction. Typical histograms of the 0.4-1k *g* and 10-22k *g* (Figure 1 F) samples show an increase in monolayer and few-layer nanosheets and a narrowing of the distribution with increasing centrifugation speed. Histograms of all other sizes are shown in the SI (Figure S1). The arithmetic mean values of nanosheet layer number, <N>, is plotted versus the central *g*-force in figure 1H and shows significant variation over the fractions from ~19 to 3.5. Experimentally, <N> followed a power-law with an exponent of -0.4.

Atomic force microscopy (AFM) can be used to measure both nanosheet thickness and lateral dimensions; this means, for each nanosheet of a given thickness, the volume can be estimated as thickness×length×width. This allows for the calculation of the volume-fraction-weighted mean layer number, $\langle N \rangle_{Vf} = \sum N^2 LW / \sum NLW$, where the summations are over all nanosheets. This is an alternative measure of nanosheet thickness which reflects the fact that mass tends to be concentrated in thicker nanosheets (the difference between <N> and <N>$_{Vf}$ is akin to the difference between number-average-molecular-weight and weight-average-molecular-weight in polymer physics).[34] We find <N>$_{Vf}$ to be directly proportional to <N> with a ratio of ~1.5 (see SI, figure S2). As a result, both can be used to express the nanosheet thickness. We have added a plot of <N>$_{Vf}$ vs. central *g*-force in figure 1H.

*Dependence of optical spectra on nanosheet dimensions*

In the case of many 2D materials, including TMDs and graphene, it has been shown that optical extinction (and absorbance) spectra change systematically with nanosheet dimensions.[17-19, 33, 35-36] In this work we use UV Vis extinction and absorbance spectroscopy to investigate the effect of nanosheet size and thickness on the optical properties of liquid-exfoliated BN. Extinction spectra of the dispersions were measured in the standard transmission mode while absorbance spectra were acquired with the sample in the centre of an integrating sphere.[17, 37] It should be noted that the extinction (Ext) is a combination of both the absorption (Abs) and scattering (Sca) where Ext(λ) = Abs(λ) + Sca(λ).[38]

Optical extinction spectra (extinction is related to the transmittance, T, via $T = 10^{-Ext}$, where $Ext = \varepsilon Cl$, with ε the extinction coefficient, C the nanosheet concentration and l as the path length) were measured for the various samples produced. Spectra are shown in figure 2A and show a peak at ~6.1 eV (205 nm). Aside from this peak, the spectra are dominated by a broad scattering background,[5, 17, 39] especially for the fractions containing larger nanosheets. Clearly the shape of this scattering background is highly dependent on nanosheet size. This is a significant problem as the extinction coefficient is usually considered to be an intrinsic property which can be used to determine dispersion concentration. It is clear that concentration measurements are only possible if the size-dependence of the extinction coefficient is determined.

The concentration of each dispersion was determined gravimetrically, i.e. by filtering a known volume of LPE BN and weighing the resultant white powder after washing with ~ 500 mL of water. This allowed us to convert *Ext* to extinction coefficient, with the data plotted at a fixed photon energy ($\varepsilon_{3.1eV}$) *versus* the mean nanosheet length, <L>, in figure 2B. We find a clear relationship between $\varepsilon_{3.1eV}$ and <L> which empirically can be described by

$$\varepsilon_{3.1eV} = 4 \times 10^{-4} \langle L \rangle^{2.55} \tag{1}$$

where <L> is in nm and $\varepsilon_{3.1eV}$ is in Lg$^{-1}$m$^{-1}$. We note that this is actually a measure of the scattering coefficient in this, non-resonant, regime. Once <L> has been measured, for example by TEM, equation 1 can be used to find the extinction coefficient appropriate to the nanosheet length under study. Then, $\varepsilon_{3.1eV}$ can be used to obtain the nanosheet concentration (using $C = Ext_{3.1eV} / l\varepsilon_{3.1eV}$). Alternatively, if the extinction coefficient is measured, this information can be used to determine <L>.

Of more basic interest than extinction is absorbance. As described recently, an integrating sphere can be used to separate the extinction spectra into their constituent absorbance and scattering components (Figure 3C).[17, 19, 40] The scattering spectra (figure 2C inset) followed power-law decays in the non-resonant regime, as described previously for dispersions of MoS$_2$ nanosheets.[17] More importantly, the absorbance spectra displayed a well-defined peak at 6.06-6.13 eV (205-202 nm) and an absorption edge at ~5.8 eV (~213 nm). In addition, the absorbance falls to zero at energies below 3.5 eV, consistent with a wide-bandgap semiconductor and confirming that the majority of the signal detected in the extinction spectra is due to scattering (SI Figure S5). However, we note that some small unexpected features were observed close to 4.2 eV. These will be discussed briefly below.

The main absorbance peak is attributed to free excitons associated with the band to band transition.[41-43] In BN grown by metal-organic chemical vapor deposition, these excitons were previously observed in photoluminescence measurements and are typically located at ~5.7 eV with impurity bound excitons also having been observed at ~5.5 eV, respectively.[41-43] In the absorbance spectra of LPE BN presented here, the excitonic peak is upshifted compared to these literature values, suggesting a Stokes-shift of up to 0.4 eV. However, the different dielectric environment of the two sample types may also result in different exciton binding energies also contributing to this shift.

The data in figure 2C shows a small shift in the exciton energies over the range of the size-selected fractions. Because the exciton binding energy is sensitive to a combination of confinement effects and dielectric screening, we would expect the excitonic peak position to vary with nanosheet thickness. This is more clearly seen in the second derivative of the peak plotted in figure 2D which clearly shows the spectra to redshift as the nanosheet size increases. The peak position, $E_{Abs}$, is plotted versus $<N>_{Vf}$ in figure 3E and falls gradually from ~6.12 eV for a mean weighted thickness of 3.4 layers to ~6.06 eV for mean weighted thickness of 27 layers. In this thickness range, the thickness–dependence of the peak position can be empirically fitted using an appropriate empirical function (dashed line). This can be rearranged to give an equation which allows us to determine $<N>_{Vf}$, once the peak absorbance is known:

$$\langle N \rangle_{Vf} = 10^{17.2(6.15-E_{Abs})} \tag{3}$$

where $E_{Abs}$ is in eV.

Magnified views of the absorption curves, focusing on the region near 4 eV are shown in figure 2H. The spectra associated with larger nanosheets clearly show well-defined features in this regime. Weak features close to 4 eV are often observed in BN absorption spectra. These are typically assigned to donor-acceptor-pair transitions involving a nitrogen vacancy donor and a deep level acceptor such as carbon atoms occupying the nitrogen vacancy site.[43-45] It is interesting to note that these impurity-related transitions decrease in intensity with decreasing layer number (Figure 2H inset). This is difficult to rationalize, as these impurities are attributed to substitutional defects in the BN lattice. Another possible explanation is that there is a zero-phonon transition at 4.15 eV with phonon replicas at higher energy.[46] Future studies using different BN starting materials are required to shine light on this phenomenon now that absorbance spectroscopy can be used as a readily available technique to not only investigate nanosheet length and thickness, but also the sample purity.

To confirm that this peak shift is a manifestation of nanosheet thickness, we calculated at different levels of theory the variation of absorption spectra with number of layers. Many-body effects as well as electron-hole interactions at the level of Bethe-Salpeter equation (BSE) are included and compared to the hybrid functional HSE06 (see *Methods* for details). Figure 3A presents the value for the energy at which the first large peak occurs in the experimental absorption spectrum (see for example Figure 2C-E) and equivalent values extracted from QS$GW$+BSE, $G_0W_0$+BSE and HSE06 simulations. Good agreement is observed between measurements and HSE06 with most of the difference in the range of few tenths of meV's. A slight overestimation of around 90 meV for QS$GW$+BSE is observed which tends to be the case when using self-consistent $GW$ approaches.[47] However, the popular $G_0W_0$+BSE approach[48] largely underestimated the measured peak position by values in the range of 0.22-0.42 eV for nanosheets. This indicates the inaccuracy of the initial DFT Hamiltonian on subsequent many-body calculations without further optimization of the electron screening self-consistently.[49-51] It is worth mentioning that such discrepancy between $G_0W_0$+BSE and measurements also appears in bulk by a larger amount ~0.47 eV, which is minimized for HSE06 and QS$GW$+BSE simulations.

A comparison between the calculated (QS$GW$+BSE) and measured absorption spectra is shown in Figure 4B. We see excellent agreement between the two, with a slight overestimation of the calculated spectrum. Such difference comes from the fact that the QS$GW$ approach is known to overestimate the value of the fundamental gap;[47] and the difference in layer number between calculation (2L) and sample thickness (3.5L). For illustrative purposes, we have included the band structure for bilayer in the inset of Figure 4B at QS$GW$ level. We show the four transitions (two valence states and two conduction states) that contribute most to the large excitonic peak in the spectrum -- just after the onset of absorption. This peak is not present at the level of the independent particle approximation (RPA), hence the justification for solving the BSE. The band structure also illustrates the calculated reduction in the gap as a result of including excitonic effects. Overall, different levels of theory capture the variation of the peak shift with layer thickness which points to a novel optimization parameter on the electronic and optical properties of supposedly inert h-BN layers.

*Dependence of Raman spectra on nanosheet dimensions*

Raman spectroscopy has evolved as a powerful tool to characterize 2D materials, as the spectra typically contain information on nanosheet thickness, defect content, strain, doping, etc.[21, 52] However, compared to graphene or other 2D materials, the Raman spectra of BN are relatively poor in information. They are dominated by a single phonon mode, the so-called G band at around 1366 cm$^{-1}$ (in addition to low frequency modes that are often not accessible[53]).[54-57] A major issue with the Raman spectroscopy on BN is that the material is not resonantly excited, so the recorded signal is very weak. In addition, only minor peak shifts have been observed.[56] For example, only the monolayer was reported to exhibit sample dependent blue-shifts compared to the bulk material with the magnitude of the shift also depending on strain.

To test whether we can nonetheless extract information, we subjected our size-selected LPE nanosheets to Raman spectroscopy after deposition on Si/SiO$_2$ wafers. The laser power was kept as low as possible to avoid heating effects (see SI, figure S6). The normalized Raman spectra are shown in Figure 4A. In addition to the G-mode (~1366 cm$^{-1}$), a secondary smaller and broad group of peaks are visible at higher Raman shift (1400-1470 cm$^{-1}$) with we attribute to sodium cholate (see SI, figure S7). In addition to this mode, we expect a sodium cholate mode at ~1365.2 cm$^{-1}$, i.e. very close to the h-BN G-mode. To eliminate the effect of this SC mode, and to determine the width and position of the h-BN G-mode as accurately as possible, we therefore fit the main Raman peak to two Lorentzians, constraining one using the known position and width of the SC mode. An example of such a fit is shown for the 5-10 k-*g* sample (Figure 4B). The resultant positions of the h-BN G-bands are plotted as a function of <N> in Figure 4C. As expected from literature, the peak position does not change systematically as function of nanosheet thickness and are centered around 1366 cm$^{-1}$ as indicated by the dashed line.

However, the G-mode does show variations in the peak width across the obtained fractions. We have analyzed the full width at half maximum ($\Gamma_{\text{G-band}}$) of the h-BN G-mode extracted from the Lorentzian fits. As shown in figure 4D, we find a near-linear scaling of the G-mode width with $1/\langle N \rangle_{Vf}$ implying that the broadening is related to the nanosheet surfaces (i.e. the basal planes).

A possible explanation of the broadening with nanosheet thickness would be related to solvatochromic effects, as the BN units in the thinner nanosheets are more completely surrounded by residual surfactant and water. To test whether the width of the G-mode is significantly influenced by the dielectric environment, a drop of *N*-cyclohexylpyrrolidone (CHP) was placed on the deposited BN nanosheets and Raman spectra acquired before and after the CHP treatment. As shown in the SI (figure S9), $\Gamma_{\text{G-band}}$ of this sample increases from ~ 9 cm$^{-1}$ to ~12 cm$^{-1}$ confirming that the nature of this broadening is solvatochromism at the outer monolayer-liquid interface.

With this in mind, we can generate a very simple model to describe the thickness dependence of $\Gamma_{\text{G-band}}$. We propose that the solvochromatic increase in linewidth compared to bulk scales with the faction of monolayer surfaces within the nanosheet which are exposed to the environment. This implies a scaling of the form ~2/(N+1) suggesting a width-thickness relationship of:

$$\Gamma_{\text{G-band}} = \Gamma_{\text{G-band}}^{\text{Bulk}} + \frac{2\Delta\Gamma_{\text{M-B}}}{\langle N \rangle_{Vf} + 1}$$

where $\Delta\Gamma_{M-B}$ is the width change going from bulk to monolayer. Applying his function to the data in figure 4D yields a very good fit and nicely captures small deviations from pure $1/\langle N\rangle_{Vf}$-type behavior. In this case, the data is consistent with $\Delta\Gamma_{M-B}=8.7$ cm$^{-1}$, higher than the value of 3-4 cm$^{-1}$, implied by the data of Gorbachev for BN nanosheets on a SiO$_2$ wafer.[56] The difference may be partly due to the fact that their nanosheets were exposed to air on one side whereas ours are likely coated with sodium cholate on both sides. In addition, the fit gives $\Gamma_{G\text{-band}}^{Bulk}=8.5$ cm$^{-1}$. Considering that a FWHM of 8 cm$^{-1}$ is found in high quality BN crystals,[53] this is very reasonable and suggests that basal plane defects (which would also broaden the BN G-mode[58]) in our LPE BN samples produced from commercial powder are minor.

Rearranging the above equation yields a relationship which allows to estimate $\langle N\rangle$ once the G-band width has been measured:

$$\langle N\rangle_{Vf} = \frac{2\Delta\Gamma_{M-B}}{\left(\Gamma_{G\text{-band}} - \Gamma_{G\text{-band}}^{Bulk}\right)} - 1 = \frac{17.2}{\left(\Gamma_{G\text{-band}} - 8.5\right)} - 1$$

where $\Gamma_{G\text{-band}}$ is in cm$^{-1}$. We note that this metric may be of limited use because the effects of solvatochromism will make the value of $\Delta\Gamma_{M-B}$ system dependent. However, we note that the thickness dependence of the Raman linewidth could be useful for measuring the thickness of CVD grown multilayer BN, where the environments at top and bottom the nanosheets are well-defined.

We note that a plot of $\Gamma_{G\text{-band}}$ versus $1/\langle L\rangle$ supports the idea that the broadening is related to nanosheet thickness, as the data in this case does not scale cleanly with $1/\langle L\rangle$ (SI, Figure S8). Note that the samples from the overnight centrifugation with a different $\langle N\rangle$-$\langle L\rangle$ relationship compared to the standard size-selected samples fall on the same curve when the G-band width is plotted as function of $1/\langle N\rangle$, but that this is not the case for $1/\langle L\rangle$.

**Conclusion**

By studying the dependence of extinction, absorption and Raman spectra on the dimensions of size-selected h-BN nanosheets, we have proposed metrics for estimating nanosheet thickness from optical spectra. The nanosheet thickness can be found from the position of the maximum in the absorbance spectrum due to thickness-dependent excitonic confinement. In addition, minor features in the absorbance spectra attributed to impurities can be used to assess the nanosheet quality. Alternatively, the nanosheet thickness can be extracted from the width of the Raman G-band due to the presence of solvatochromic effects. We suggest that both thickness measurements could be applied to BN multilayers grown by methods such as CVD.

**Methods**

*Sample Preparation*

BN dispersions were prepared by probe sonicating (VibraCell CVX, 750W) powder (Sigma Aldrich ~ 1 µm, 98%) at a concentration of 30 g $L^{-1}$ dispersed in a 6 g $L^{-1}$ aqueous solution of sodium cholate (Sigma Aldrich BioXtra, ≥99%) for 1 hr at 60% amplitude. The dispersion was then centrifuged in a Hettich Mikro 220R centrifuge equipped with a fixed-angle rotor 1016 at 2260 $g$ for 2 hrs. The supernatant was removed and the sediment was redispersed in fresh surfactant solution (conc=2 g $L^{-1}$) and subsequently sonicated for 6 hrs at 60% amplitude with a pulse of 6 on and 2 off. The resultant stock dispersion was centrifuged at 27 $g$ for 2 h, sediment discarded and the supernatant subjected to size selection. For the size selection of nanosheets, we used a centrifugation cascade increasing the speed and moving the supernatant on to the next stage each time. The sediment after each centrifugation was collected and redispersed in fresh surfactant solution. The speeds used were 0.1k $g$, 0.4k $g$, 1k $g$, 5k $g$, 10k $g$, 22k $g$. For centrifugation < 3k $g$, a Hettich Mikro 220R centrifuge equipped with a fixed-angle rotor 1016 (50 mL vials filled with 20 mL each). For centrifugation > 3k $g$, a Beckman Coulter Avanti XP centrifuge was used with a JA25.15 rotor with 14 mL vials (Beckman Coulter), filled with 10 mL dispersion each. All centrifugation was performed for 2 h at 15°C. The data in Figure 1 uses the central g-force to express the consecutive centrifugation speeds. The central g force for a 0.4-1k $g$ trapping (supernatant from 0.4k $g$ then centrifuged at 1k $g$ with the sediment collected) for example is 0.7k $g$. In addition to these samples from the standard cascade, two additional samples were prepared with the goal to achieve a different quantitative relationship between lateral size and layer number. For this purpose, the samples 0.1-0.4k g and 0.4-1k g were centrifuged for 16 h at 50 g (Hettich Mikro 220R centrifuge, fixed angle rotor 1195-A, 1.5 mL vials) The concentration of BN in the fractions was determined by filtration and weighing (alumina membranes pore size 0.02 µm). Prior to weighing, the samples were washed with 600 mL of deionised water and dried in vacuum at 70°C.

*Characterization*

Optical extinction and absorbance measurements were carried out on a Cary 6000i spectrometer in quartz cuvettes. The spectrometer was fitted with an integrating sphere for absorbance measurements. In this case, the cuvettes were placed in the center of the sphere and the absorbance was measured with 10 cm$^{-1}$ increments and a band widths of 2 nm. The optical density of the BN in the absorbance measurement was adjusted to 0.3-0.4 at the peak. The measurements of both extinction and absorbance spectra allows for the calculation of scattering spectra (Ext-Abs). In this case, the spectra were measured with 0.5 nm increments to give a higher resolution at lower energy. A Bruker Icon Dimension Atomic Force microscope in ScanAsyst mode with Bruker Oltespa-R3 cantilevers was used for AFM measurements. Each liquid dispersion (10 µL) was diluted until the sample was transparent drop cast onto a preheated (180 °C) Si/SiO$_2$ (300 nm oxide layer) wafer and individually deposited nanosheets analyzed. To correct the nanosheet length due to tip broadening, we used a previously established length correction.[18] Raman spectroscopy was carried out on a Renishaw InVia-Reflex Confocal Raman Microscope with a 532 nm excitation laser in air under ambient conditions. The Raman emission was collected by a 50×, long working distance objective lens in streamline mode and dispersed by a 2400 l/mm grating with 10 % of the laser power (<1.4 mW). Liquid dispersions were dropped (∼ 20 µL) onto Si/SiO$_2$ wafers and left to dry in air before measuring. Minimum 5 spectra on different positions were recorded and averaged. In the streamline mode, where a larger sample area is sampled we did not observe spot to spot variations except for absolute intensities.

*Ab initio density functional theory calculations and many-body perturbation approaches*

DFT calculations were performed for layered *h*-BN with 1-19 layers plus the bulk system using the VASP code.[59] The unit cells were set up with AA' stacking and lattice constants: a=b=2.5 Å, c=3.32 Å. The vacuum space used due to the periodic boundary conditions was always beyond 20 Å for each system. The *1s* states in both B and N were treated as part of the core within a PAW pseudopotential[60] and a plane-wave cut-off of 800eV was used throughout. The cells were relaxed with a $\Gamma$-centered 6x6xN (*N*=1 for layered materials and *N*=2 for bulk) *k*-mesh and using the state of the art hybrid HSE06 functional[61-63] with the

Tkatchenko and Scheffler[64] method to account for van der Waals interactions. The interlayer distance varied between 3.32 and 3.35 Å as a result of the relaxation.

To correct any limitations observed at the level of HSE06 method, we have performed simulations for the optical properties at the level of many-body *GW* approximations plus Bethe-Salpeter equation (BSE). It has been shown, however, that $G_0W_0$ (the single-shot method[48]) built from the LDA or GGA Hamiltonian significantly underestimates the fundamental band gap in certain systems, of which *h*-BN is one.[49-51] Therefore, to calculate the electronic and optical properties we have used Quasi-Particle Self-Consistent GW (QS*GW*) method,[47, 65-66] whereby the optimum starting Hamiltonian is determined using the *GW* approximation iteratively. This optimum starting point is determined by minimizing the perturbation between the self-energy $\Sigma$ and the exchange-correlation potential $V_{XC}$, as in $\Sigma - V_{XC}$. The QS*GW* method tends to overestimate band gaps due mainly to the fact that vertex corrections are missing[47]. We include excitonic effects in this work by solving the BSE equation for the polarization.[67-69] The macroscopic dielectric function obtained within the QS*GW*+BSE method is

(1)
$$\epsilon(\omega) = 1 - \lim_{q \to 0} \frac{8\pi}{|q|^2 \Omega N_k N_\sigma} \times \sum_{n_1 n_2 k n_3 n_4 k'} (f_{n_4 k'+q} - f_{n_3 k'}) \rho_{n_1 n_2 k}(q) [H(q) - \omega]^{-1}_{n_1 n_2 k n_3 n_4 k'} \rho^*_{n_3 n_4 k'}(q)$$

where $\Omega$, $N_k$ and $N_\sigma$ are the cell volume, number of *k*-points in the full Brillouin zone and number of spin channels treated explicitly and $f_{nk}$ are the QS*GW* single-particle occupations. The transition dipole matrix elements (often referred to as oscillators) are

(2) $$\rho_{n_1 n_2 k}(q) = \langle \psi_{n_2 k+q} | e^{iq \cdot r} | \psi_{n_1 k} \rangle$$

where the eigenfunctions are the QS*GW* ones. Finally, the effective two-particle Hamiltonian, *H*, in Eq. 1 is

(3) $$H_{n_1 n_2 k n_3 n_4 k'} = (\varepsilon_{n_2 k+q} - \varepsilon_{n_1 k}) \delta_{n_1 n_3} \delta_{n_2 n_4} \delta_{kk'} - (f_{n_2 k+q} - f_{n_1 k}) K_{n_1 n_2 k n_3 n_4 k'}(q)$$

with $\varepsilon_n$ the QSGW single-particle eigenenergies. The kernel is $K=2V-W$, with $V$ the bare Coulomb interaction and W the screened Coulomb interaction. If we set $K=0$ then we are at the level of the independent particle approximation also known as the random phase approximation (RPA). The imaginary part of the macroscopic dielectric function, $\varepsilon_2$, then produces the theoretical absorption spectrum which can be directly compared with the experimental spectrum. The structure for the systems in the QSGW calculation were the VASP relaxed structures discussed above. Due to the large memory requirements of the BSE we only considered states that are within ±6eV (at $\Gamma$) of the Fermi level when constructing $H$ (Eq. 3). Transitions not included in $H$ contribute to $\varepsilon_M$ at the level of independent particle transitions (only the first term in Eq. 3 present). We assume that the kernel in Eq. 3 is static and adopt the Tamm-Dancoff approximation,[70] whereby we neglect the coupling between positive and negative energy transitions.

**Figures**

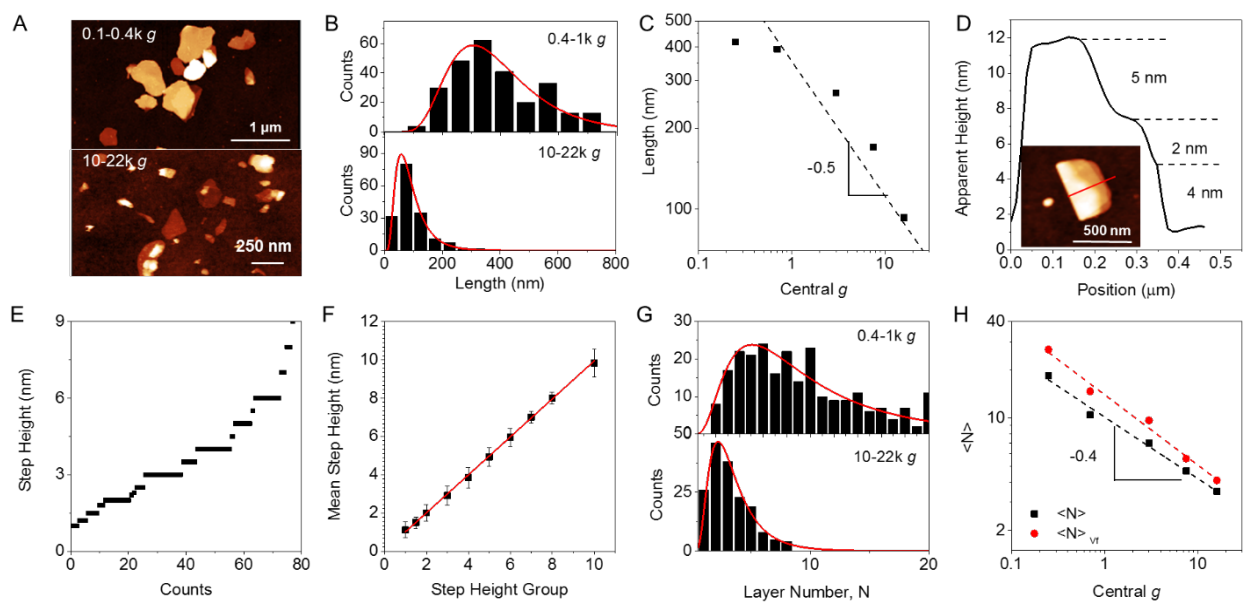

**Figure 1**: Microscopic characterization of size selected BN nanosheets. **(A)** Representative images of BN nanosheets from the 0.1-0.4k *g* and 10-22k *g* fractions. **(B)** Histograms of nanosheet lateral size for the fractions shown in A. **(C)** Mean nanosheet length as a function of central centrifugal force *g*. **(D)** Height profile along the line of the nanosheet in the inset showing clear, resolvable steps each consisting of multiple monolayers. **(E)** Step heights of >70 BN nanosheets in ascending order. The step height clustered in groups and is always found to be a multiple of ~1 nm, which is the apparent height of one monolayer. **(F)** The mean height for each group (the error is the sum of the mean step height error and the standard deviation in step height within a given group) is plotted in ascending order with the slope giving a mean monolayer step height of 0.99 ± 0.01 nm. **(G)** Histograms of layer number, N, for the fractions shown in A. **(H)** Mean layer number <N> and volume-fraction-weighted mean layer number <N> Vf weighted, both plotted versus central centrifugal force *g*.

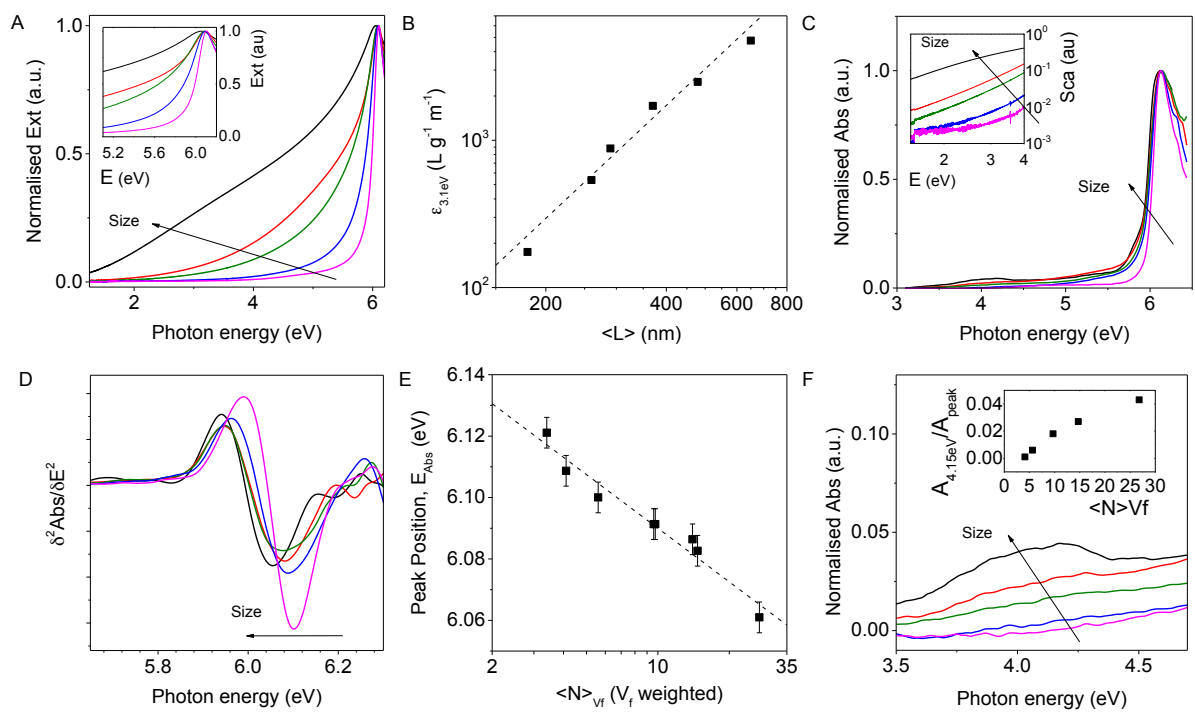

**Figure 2:** Extinction and absorbance spectroscopy of BN nanosheets. **(A)** Optical extinction spectra normalized to peak maxima showing dependence on nanosheet size. Inset: magnified view of peak region. **(B)** Extinction coefficient at 3.1 eV, $\varepsilon_{3.1eV}$, plotted as a function of nanosheet mean length <L> as measured by AFM. **(C)** Normalised optical absorption spectra for different nanosheet sizes. Inset: Scattering spectra in non-resonant regime. **(D)** Second derivative of the peak region of the absorption spectra. **(E)** Peak position of the absorbance spectra plotted versus the volume fraction weighted average layer number <N> Vf weighted as measured by AFM. **(F)** Magnified view of absorption spectra in the energy range close to 4 eV. Inset: Absorbance at 4.15 eV normalized to peak absorbance plotted versus mean nanosheet thickness.

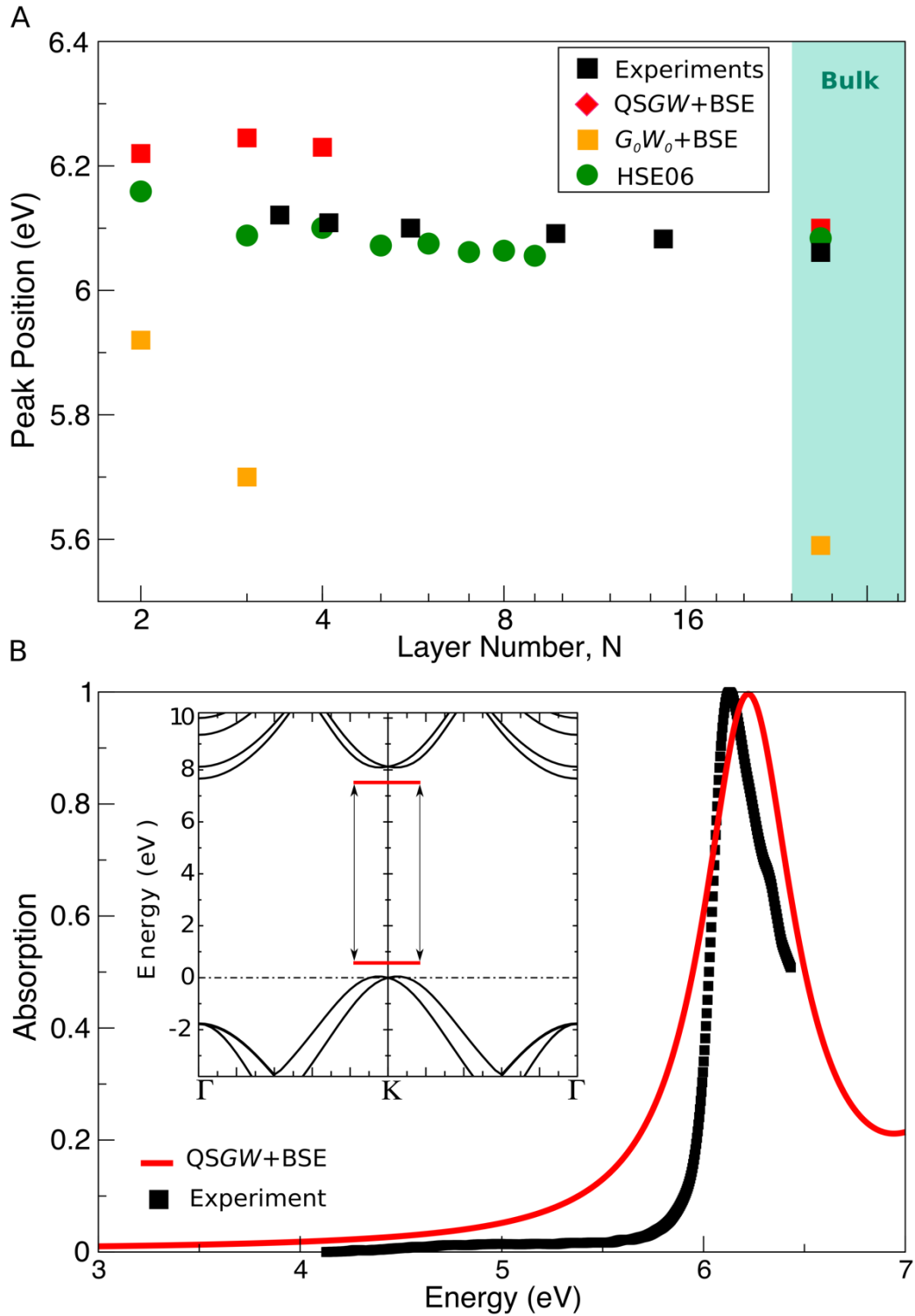

**Figure 3:** *Ab initio* calculations: Many-body perturbation approaches and hybrid functional. **(A)** Peak position extracted from absorption spectra versus number of h-BN layers measured experimentally, black squares, and compared to calculations at different levels of theory: quasiparticle self-consistency GW plus BSE, QS$GW$+BSE (red diamonds); single-shot GW plus BSE, $G_0W_0$+BSE (orange squares); and hybrid functional, HSE06 (green circles).

Comparison to bulk results is performed at the faint green zone. (**B**) Optical absorption spectrum for the sample with 3.5 layers (black dots) and the macroscopic dielectric function calculated at the level of QSGW+BSE for bilayer (solid line). A numerical broadening of 272 meV was used to plot the calculated absorption, which accounts for the different profile between both spectra even though the difference at peak position is by less than 80 meV. The inset presents the band structure for bilayer along the high symmetry points $\Gamma - K - \Gamma$ indicated and the red lines represent the effect of including excitonic effects through solving the Bethe-Salpeter equation. The two bands at the valence band maximum and two bands at the conduction band minimum couple and are largely responsible for the large excitonic peak in the absorption spectrum.

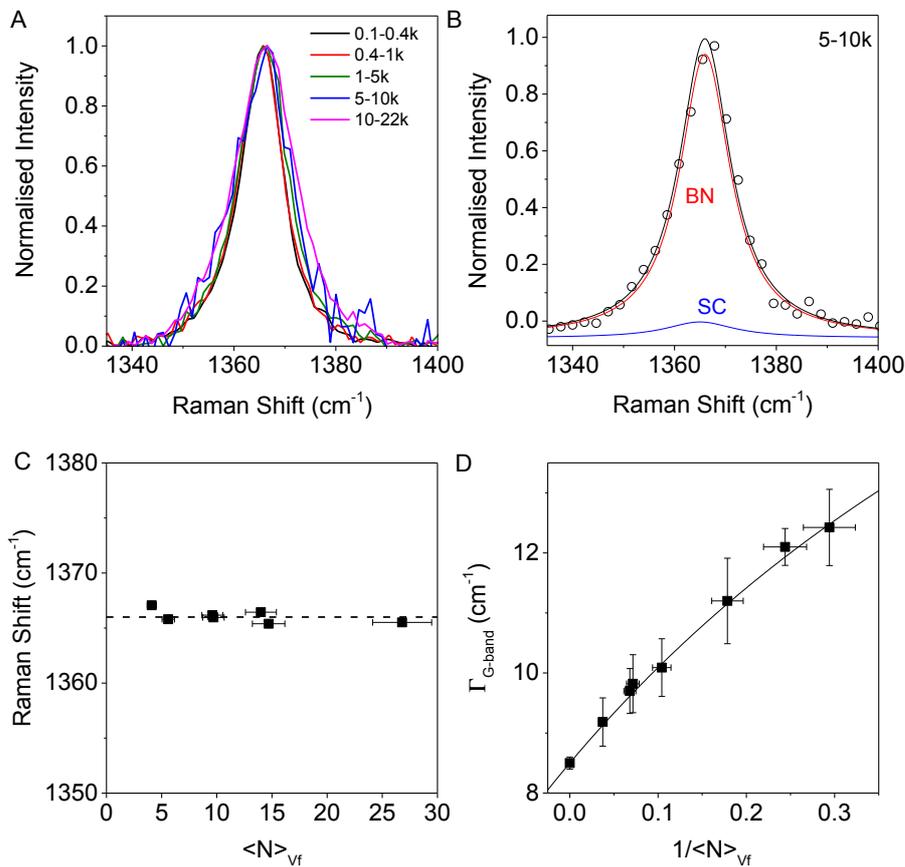

**Figure 4:** Raman spectroscopy of size-selected BN nanosheets. (**A**) Raman spectra of size-selected BN nanosheet dispersions normalized to the peak maxima at the G band frequency (~1366 cm$^{-1}$). (**B**) Fitted Raman spectrum of the fraction 5-10k *g*, normalized to the maximum intensity and fitted to two lines, one representing h-BN and the other representing sodium cholate. (**C**) Plot of the h-BN G-band position as function of mean layer number. The

G-band is centered at 1367 cm$^{-1}$ (dashed line). **(D)** h-BN G-band peak width (full width and half maximum, FWHM, from fit) as function of the inverse nanosheet thickness. The dashed line is a fit to equation * consistent with broadening being due to solvatochromic effects.

## Acknowledgement

The research leading to these results has received funding from the European Union's Horizon 2020 under grant agreement n°604391 Graphene Flagship. C.B. acknowledges the German research foundation DFG under Emmy-Noether grant BA4856/2-1. We thank Jana Zaumseil for access to the infrastructure at the Chair of Applied Physical Chemistry, Heidelberg. C.J.S. and T.T. are grateful for financial support from ETH startup funding. D.S. thanks his EPSRC studentship. B.C. and M.G. are grateful for support from the Engineering and Physical Sciences Research Council, under grant EP/M011631/1. E.J.G.S. acknowledges the use of computational resources from the UK national high-performance computing service (ARCHER) for which access was obtained via the UKCP consortium (EPSRC grant ref EP/K013564/1); the UK Materials and Molecular Modelling Hub for access to THOMAS supercluster, which is partially funded by EPSRC (EP/P020194/1). The Queen's Fellow Award through the grant number M8407MPH, the Enabling Fund (A5047TSL), and the Department for the Economy (USI 097) are also acknowledged.

## Additional information

Supplementary information is available.

## Competing financial interests

The authors declare no competing financial interests.

## References

1. Golberg, D.; Bando, Y.; Huang, Y.; Terao, T.; Mitome, M.; Tang, C. C.; Zhi, C. Y., Boron Nitride Nanotubes and Nanosheets. *Acs Nano* **2010,** *4* (6), 2979-2993.

2.      Bonaccorso, F.; Lombardo, A.; Hasan, T.; Sun, Z. P.; Colombo, L.; Ferrari, A. C., Production and processing of graphene and 2d crystals. *Mater. Today* **2012,** *15* (12), 564-589.
3.      Hemmi, A.; Bernard, C.; Cun, H.; Roth, S.; Klockner, M.; Kalin, T.; Weinl, M.; Gsell, S.; Schreck, M.; Osterwalder, J.; Greber, T., High quality single atomic layer deposition of hexagonal boron nitride on single crystalline Rh(111) four-inch wafers. *Review of Scientific Instruments* **2014,** *85* (3), 4.
4.      Gannett, W.; Regan, W.; Watanabe, K.; Taniguchi, T.; Crommie, M. F.; Zettl, A., Boron nitride substrates for high mobility chemical vapor deposited graphene. *Applied Physics Letters* **2011,** *98* (24), 3.
5.      Coleman, J. N.; Lotya, M.; O'Neill, A.; Bergin, S. D.; King, P. J.; Khan, U.; Young, K.; Gaucher, A.; De, S.; Smith, R. J.; Shvets, I. V.; Arora, S. K.; Stanton, G.; Kim, H. Y.; Lee, K.; Kim, G. T.; Duesberg, G. S.; Hallam, T.; Boland, J. J.; Wang, J. J.; Donegan, J. F.; Grunlan, J. C.; Moriarty, G.; Shmeliov, A.; Nicholls, R. J.; Perkins, J. M.; Grieveson, E. M.; Theuwissen, K.; McComb, D. W.; Nellist, P. D.; Nicolosi, V., Two-Dimensional Nanosheets Produced by Liquid Exfoliation of Layered Materials. *Science* **2011,** *331* (6017), 568-571.
6.      Zhi, C. Y.; Bando, Y.; Tang, C. C.; Kuwahara, H.; Golberg, D., Large-Scale Fabrication of Boron Nitride Nanosheets and Their Utilization in Polymeric Composites with Improved Thermal and Mechanical Properties. *Advanced Materials* **2009,** *21* (28), 2889-+.
7.      Cui, X.; Lee, G. H.; Kim, Y. D.; Arefe, G.; Huang, P. Y.; Lee, C. H.; Chenet, D. A.; Zhang, X.; Wang, L.; Ye, F.; Pizzocchero, F.; Jessen, B. S.; Watanabe, K.; Taniguchi, T.; Muller, D. A.; Low, T.; Kim, P.; Hone, J., Multi-terminal transport measurements of MoS2 using a van der Waals heterostructure device platform. *Nature Nanotechnology* **2015,** *10* (6), 534-540.
8.      Dean, C. R.; Young, A. F.; Meric, I.; Lee, C.; Wang, L.; Sorgenfrei, S.; Watanabe, K.; Taniguchi, T.; Kim, P.; Shepard, K. L.; Hone, J., Boron nitride substrates for high-quality graphene electronics. *Nature Nanotechnology* **2010,** *5* (10), 722-726.
9.      Roth, S.; Matsui, F.; Greber, T.; Osterwalder, J., Chemical Vapor Deposition and Characterization of Aligned and Incommensurate Graphene/Hexagonal Boron Nitride Heterostack on Cu(111). *Nano Letters* **2013,** *13* (6), 2668-2675.
10.     Palacios-Berraquero, C.; Barbone, M.; Kara, D. M.; Chen, X. L.; Goykhman, I.; Yoon, D.; Ott, A. K.; Beitner, J.; Watanabe, K.; Taniguchi, T.; Ferrari, A. C.; Atature, M., Atomically thin quantum light-emitting diodes. *Nat. Commun.* **2016,** *7*, 6.
11.     Xie, S. B.; Istrate, O. M.; May, P.; Barwich, S.; Bell, A. P.; Khana, U.; Coleman, J. N., Boron nitride nanosheets as barrier enhancing fillers in melt processed composites. *Nanoscale* **2015,** *7* (10), 4443-4450.
12.     Taha-Tijerina, J.; Narayanan, T. N.; Gao, G. H.; Rohde, M.; Tsentalovich, D. A.; Pasquali, M.; Ajayan, P. M., Electrically Insulating Thermal Nano-Oils Using 2D Fillers. *Acs Nano* **2012,** *6* (2), 1214-1220.
13.     Kelly, A. G.; Finn, D.; Harvey, A.; Hallam, T.; Coleman, J. N., All-printed capacitors from graphene-BN-graphene nanosheet heterostructures. *Applied Physics Letters* **2016,** *109* (2).
14.     Yang, H. F.; Withers, F.; Gebremedhn, E.; Lewis, E.; Britnell, L.; Felten, A.; Palermo, V.; Haigh, S.; Beljonne, D.; Casiraghi, C., Dielectric nanosheets made by liquid-phase exfoliation in water and their use in graphene-based electronics. *2d Materials* **2014,** *1* (1).
15.     Zhu, J.; Kang, J.; Kang, J.; Jariwala, D.; Wood, J. D.; Seo, J.-W. T.; Chen, K.-S.; Marks, T. J.; Hersam, M. C., Solution-Processed Dielectrics Based on Thickness-Sorted Two-Dimensional Hexagonal Boron Nitride Nanosheets. *Nano Letters* **2015,** *15* (10), 7029-7036.
16.     Kelly, A. G.; Hallam, T.; Backes, C.; Harvey, A.; Esmaeily, A. S.; Godwin, I.; Coelho, J.; Nicolosi, V.; Lauth, J.; Kulkarni, A.; Kinge, S.; Siebbeles, L. D. A.; Duesberg, G. S.; Coleman, J. N., All-printed thin-film transistors from networks of liquid-exfoliated nanosheets. *Science* **2017,** *356* (6333), 69-72.
17.     Backes, C.; Smith, R. J.; McEvoy, N.; Berner, N. C.; McCloskey, D.; Nerl, H. C.; O'Neill, A.; King, P. J.; Higgins, T.; Hanlon, D.; Scheuschner, N.; Maultzsch, J.; Houben, L.; Duesberg, G. S.; Donegan, J. F.; Nicolosi, V.; Coleman, J. N., Edge and Confinement Effects Allow in situ Measurement of Size and Thickness of Liquid-Exfoliated Nanosheets. *Nat. Commun.* **2014,** *5*, 4576.
18.     Backes, C.; Szydłowska, B. M.; Harvey, A.; Yuan, S.; Vega-Mayoral, V.; Davies, B. R.; Zhao, P.-l.; Hanlon, D.; Santos, E. J. G.; Katsnelson, M. I.; Blau, W. J.; Gadermaier, C.; Coleman, J.

N., Production of Highly Monolayer Enriched Dispersions of Liquid-Exfoliated Nanosheets by Liquid Cascade Centrifugation. *ACS Nano* **2016,** *10* (1), 1589-1601.
19.  Backes, C.; Paton, K. R.; Hanlon, D.; Yuan, S.; Katsnelson, M. I.; Houston, J.; Smith, R. J.; McCloskey, D.; Donegan, J. F.; Coleman, J. N., Spectroscopic metrics allow in situ measurement of mean size and thickness of liquid-exfoliated few-layer graphene nanosheets. *Nanoscale* **2016,** *8* (7), 4311-4323.
20.  Lee, C.; Yan, H.; Brus, L. E.; Heinz, T. F.; Hone, J.; Ryu, S., Anomalous Lattice Vibrations of Single- and Few-Layer MoS2. *Acs Nano* **2010,** *4* (5), 2695-2700.
21.  Ferrari, A. C.; Basko, D. M., Raman spectroscopy as a versatile tool for studying the properties of graphene. *Nat Nano* **2013,** *8* (4), 235-246.
22.  Schue, L.; Berini, B.; Betz, A. C.; Placais, B.; Ducastelle, F.; Barjon, J.; Loiseau, A., Dimensionality effects on the luminescence properties of hBN. *Nanoscale* **2016,** *8* (13), 6986-6993.
23.  Bonaccorso, F.; Bartolotta, A.; Coleman, J. N.; Backes, C., 2D-Crystal-Based Functional Inks. *Adv. Mater.* **2016,** *28* (29), 6136-6166.
24.  Niu, L. Y.; Coleman, J. N.; Zhang, H.; Shin, H.; Chhowalla, M.; Zheng, Z. J., Production of Two-Dimensional Nanomaterials via Liquid-Based Direct Exfoliation. *Small* **2016,** *12* (3), 272-293.
25.  Habib, T.; Sundaravadivelu Devarajan, D.; Khabaz, F.; Parviz, D.; Achee, T. C.; Khare, R.; Green, M. J., Cosolvents as Liquid Surfactants for Boron Nitride Nanosheet (BNNS) Dispersions. *Langmuir* **2016,** *32* (44), 11591-11599.
26.  Joseph, A. M.; Nagendra, B.; Gowd, E. B.; Surendran, K. P., Screen-Printable Electronic Ink of Ultrathin Boron Nitride Nanosheets. *Acs Omega* **2016,** *1* (6), 1220-1228.
27.  Shang, J. Q.; Xue, F.; Fan, C. J.; Ding, E. Y., Preparation of few layers hexagonal boron nitride nanosheets via high-pressure homogenization. *Materials Letters* **2016,** *181*, 144-147.
28.  Yuan, F.; Jiao, W. C.; Yang, F.; Liu, W. B.; Liu, J. Y.; Xu, Z. H.; Wang, R. G., Scalable exfoliation for large-size boron nitride nanosheets by low temperature thermal expansion-assisted ultrasonic exfoliation. *Journal of Materials Chemistry C* **2017,** *5* (25), 6359-6368.
29.  Zhang, B. W.; Wu, Q.; Yu, H. T.; Bulin, C. K.; Sun, H.; Li, R. H.; Ge, X.; Xing, R. G., High-Efficient Liquid Exfoliation of Boron Nitride Nanosheets Using Aqueous Solution of Alkanolamine. *Nanoscale Research Letters* **2017,** *12*.
30.  Green, A. A.; Hersam, M. C., Solution Phase Production of Graphene with Controlled Thickness via Density Differentiation. *Nano Letters* **2009,** *9* (12), 4031-4036.
31.  Kang, J.; Sangwan, V. K.; Wood, J. D.; Hersam, M. C., Solution-Based Processing of Monodisperse Two-Dimensional Nanomaterials. *Acc. Chem. Res.* **2017,** *50* (4), 943-951.
32.  Paton, K. R.; Varrla, E.; Backes, C.; Smith, R. J.; Khan, U.; O'Neill, A.; Boland, C.; Lotya, M.; Istrate, O. M.; King, P.; Higgins, T.; Barwich, S.; May, P.; Puczkarski, P.; Ahmed, I.; Moebius, M.; Pettersson, H.; Long, E.; Coelho, J.; O'Brien, S. E.; McGuire, E. K.; Sanchez, B. M.; Duesberg, G. S.; McEvoy, N.; Pennycook, T. J.; Downing, C.; Crossley, A.; Nicolosi, V.; Coleman, J. N., Scalable production of large quantities of defect-free few-layer graphene by shear exfoliation in liquids. *Nature Materials* **2014,** *13* (6), 624-630.
33.  Hanlon, D.; Backes, C.; Doherty, E.; Cucinotta, C. S.; Berner, N. C.; Boland, C.; Lee, K.; Lynch, P.; Gholamvand, Z.; Harvey, A.; Zhang, S.; Wang, K.; Moynihan, G.; Pokle, A.; Ramasse, Q. M.; McEvoy, N.; Blau, W. J.; Wang, J.; Abellan, G.; Hauke, F.; Hirsch, A.; Sanvito, S.; O'Regan, D. D.; Duesberg, G. S.; Nicolosi, V.; Coleman, J. N., Liquid Exfoliation of Solvent-Stabilised Few-Layer Black Phosphorus for Applications Beyond Electronics. *Nat. Commun.* **2015,** *6*, 8563.
34.  Michael Rubinstein; Colby, R. H., *Polymer Physics*. Oxford University Press: 2003.
35.  Harvey, A.; Backes, C.; Gholamvand, Z.; Hanlon, D.; McAteer, D.; Nerl, H. C.; McGuire, E.; Seral-Ascaso, A.; Ramasse, Q. M.; McEvoy, N.; Winters, S.; Berner, N. C.; McCloskey, D.; Donegan, J.; Duesberg, G.; Nicolosi, V.; Coleman, J. N., Preparation of Gallium Sulfide Nanosheets by Liquid Exfoliation and Their Application As Hydrogen Evolution Catalysts. *Chemistry of Materials* **2015,** *27* (9), 3483–3493.
36.  Mandeep, S.; Enrico Della, G.; Taimur, A.; Sumeet, W.; Rajesh, R.; Joel van, E.; Edwin, M.; Vipul, B., Soft exfoliation of 2D SnO with size-dependent optical properties. *2D Materials* **2017,** *4* (2), 025110.

37. Yadgarov, L.; Choi, C. L.; Sedova, A.; Cohen, A.; Rosentsveig, R.; Bar-Elli, O.; Oron, D.; Dai, H. J.; Tenne, R., Dependence of the Absorption and Optical Surface Plasmon Scattering of MoS2 Nanoparticles on Aspect Ratio, Size, and Media. *Acs Nano* **2014,** *8* (4), 3575-3583.
38. van de Hulst, H. C., *Light Scattering by Small Particles*. Courier Corporation: 1981.
39. O'Brien, S. A.; Harvey, A.; Griffin, A.; Donnelly, T.; Mulcahy, D.; Coleman, J. N.; Donegan, J. F.; McCloskey, D., Light scattering and random lasing in aqueous suspensions of hexagonal boron nitride nanoflakes. *Nanotechnology* **2017,** *28* (47).
40. Yadgarov, L.; Choi, C. L.; Sedova, A.; Cohen, A.; Rosentsveig, R.; Bar-Elli, O.; Oron, D.; Dai, H.; Tenne, R., Dependence of the Absorption and Optical Surface Plasmon Scattering of MoS2 Nanoparticles on Aspect Ratio, Size, and Media. *ACS Nano* **2014,** *8* (4), 3575-3583.
41. Du, X. Z.; Li, J.; Lin, J. Y.; Jiang, H. X., The origins of near band-edge transitions in hexagonal boron nitride epilayers. *Applied Physics Letters* **2016,** *108* (5), 052106.
42. Doan, T. C.; Li, J.; Lin, J. Y.; Jiang, H. X., Bandgap and exciton binding energies of hexagonal boron nitride probed by photocurrent excitation spectroscopy. *Applied Physics Letters* **2016,** *109* (12), 122101.
43. Museur, L.; Kanaev, A., Near band-gap photoluminescence properties of hexagonal boron nitride. *Journal of Applied Physics* **2008,** *103* (10), 103520.
44. Serrano, J.; Bosak, A.; Arenal, R.; Krisch, M.; Watanabe, K.; Taniguchi, T.; Kanda, H.; Rubio, A.; Wirtz, L., Vibrational Properties of Hexagonal Boron Nitride: Inelastic X-Ray Scattering and Ab Initio Calculations. *Physical Review Letters* **2007,** *98* (9), 095503.
45. Wu, J.; Han, W.-Q.; Walukiewicz, W.; Ager, J. W.; Shan, W.; Haller, E. E.; Zettl, A., Raman Spectroscopy and Time-Resolved Photoluminescence of BN and BxCyNz Nanotubes. *Nano Letters* **2004,** *4* (4), 647-650.
46. Silly, M. G.; Jaffrennou, P.; Barjon, J.; Lauret, J. S.; Ducastelle, F.; Loiseau, A.; Obraztsova, E.; Attal-Tretout, B.; Rosencher, E., Luminescence properties of hexagonal boron nitride: Cathodoluminescence and photoluminescence spectroscopy measurements. *Physical Review B* **2007,** *75* (8).
47. Faleev, S. V.; van Schilfgaarde, M.; Kotani, T., All-electron self-consistent GW approximation: Application to Si, MnO, and NiO. *Physical Review Letters* **2004,** *93* (12).
48. Hybertsen, M. S.; Louie, S. G., Electron Correlation in Semiconductors and Insulators - Band-Gaps and Quasi-Particle Energies. *Physical Review B* **1986,** *34* (8), 5390-5413.
49. Berseneva, N.; Gulans, A.; Krasheninnikov, A. V.; Nieminen, R. M., Electronic structure of boron nitride sheets doped with carbon from first-principles calculations. *Physical Review B* **2013,** *87* (3).
50. Wirtz, L.; Marini, A.; Rubio, A., Optical absorption of hexagonal boron nitride and BN nanotubes. In *Electronic Properties of Novel Nanostructures*, Kuzmany, H.; Fink, J.; Mehring, M.; Roth, S., Eds. 2005; Vol. 786, pp 391-395.
51. Wirtz, L.; Marini, A.; Grüning, M.; Rubio, A. eprint arXiv:cond-mat/0508421 2005.
52. Pimenta, M. A.; del Corro, E.; Carvalho, B. R.; Fantini, C.; Malard, L. M., Comparative Study of Raman Spectroscopy in Graphene and MoS2-type Transition Metal Dichalcogenides. *Accounts of Chemical Research* **2015,** *48* (1), 41-47.
53. Stenger, I.; Schué, L.; Boukhicha, M.; Berini, B.; Plaçais, B.; Loiseau, A.; Barjon, J., Low frequency Raman spectroscopy of few-atomic-layer thick hBN crystals. *2D Materials* **2017,** *4* (3), 031003.
54. Reich, S.; Ferrari, A. C.; Arenal, R.; Loiseau, A.; Bello, I.; Robertson, J., Resonant Raman scattering in cubic and hexagonal boron nitride. *Physical Review B* **2005,** *71* (20), 205201.
55. Cai, Q.; Scullion, D.; Falin, A.; Watanabe, K.; Taniguchi, T.; Chen, Y.; Santos, E. J. G.; Li, L. H., Raman signature and phonon dispersion of atomically thin boron nitride. *Nanoscale* **2017,** *9* (9), 3059-3067.
56. Gorbachev, R. V.; Riaz, I.; Nair, R. R.; Jalil, R.; Britnell, L.; Belle, B. D.; Hill, E. W.; Novoselov, K. S.; Watanabe, K.; Taniguchi, T.; Geim, A. K.; Blake, P., Hunting for Monolayer Boron Nitride: Optical and Raman Signatures. *Small* **2011,** *7* (4), 465-468.
57. Léonard, S.; Ingrid, S.; Frédéric, F.; Annick, L.; Julien, B., Characterization methods dedicated to nanometer-thick hBN layers. *2D Materials* **2017,** *4* (1), 015028.

58. Sainsbury, T.; Satti, A.; May, P.; Wang, Z.; McGovern, I.; Gun'ko, Y. K.; Coleman, J., Oxygen Radical Functionalization of Boron Nitride Nanosheets. *Journal of the American Chemical Society* **2012,** *134* (45), 18758-18771.
59. Kresse, G.; Hafner, J., Abinitio Molecular-Dynamics for Liquid-Metals. *Physical Review B* **1993,** *47* (1), 558-561.
60. Kresse, G.; Joubert, D., From ultrasoft pseudopotentials to the projector augmented-wave method. *Physical Review B* **1999,** *59* (3), 1758-1775.
61. Heyd, J.; Scuseria, G. E.; Ernzerhof, M., Hybrid functionals based on a screened Coulomb potential. *Journal of Chemical Physics* **2003,** *118* (18), 8207-8215.
62. Heyd, J.; Scuseria, G. E.; Ernzerhof, M., Hybrid functionals based on a screened Coulomb potential (vol 118, pg 8207, 2003). *Journal of Chemical Physics* **2006,** *124* (21).
63. Krukau, A. V.; Vydrov, O. A.; Izmaylov, A. F.; Scuseria, G. E., Influence of the exchange screening parameter on the performance of screened hybrid functionals. *Journal of Chemical Physics* **2006,** *125* (22).
64. Tkatchenko, A.; Scheffler, M., Accurate Molecular Van Der Waals Interactions from Ground-State Electron Density and Free-Atom Reference Data. *Physical Review Letters* **2009,** *102* (7).
65. van Schilfgaarde, M.; Kotani, T.; Faleev, S., Quasiparticle self-consistent GW theory. *Physical Review Letters* **2006,** *96* (22).
66. "Questaal code website," https://www.questaal.org, accessed: 2017-07-04.
67. Onida, G.; Reining, L.; Rubio, A., Electronic excitations: density-functional versus many-body Green's-function approaches. *Reviews of Modern Physics* **2002,** *74* (2), 601-659.
68. Salpeter, E. E.; Bethe, H. A., A Relativistic Equation for Bound-State Problems. *Physical Review* **1951,** *84* (6), 1232-1242.
69. Kotani, T.; van Schilfgaarde, M.; Faleev, S. V., Quasiparticle self-consistent $GW$ method: A basis for the independent-particle approximation. *Physical Review B* **2007,** *76* (16), 165106.
70. Gruning, M.; Marini, A.; Gonze, X., Exciton-Plasmon States in Nanoscale Materials: Breakdown of the Tamm-Dancoff Approximation. *Nano Letters* **2009,** *9* (8), 2820-2824.